\documentclass[aps,prl,superscriptaddress,preprintnumbers,twocolumn,groupedaddress,showpacs,nofootinbib]{revtex4-1}

\pdfoutput=1

\usepackage{amssymb,amsmath}
\usepackage{graphicx}
\usepackage{url,booktabs}
\usepackage{color}

\newcommand{\met}{\not\!\!E_T}

\begin{document}

\preprint{ZU-TH 20/12, LPN12-094}
\title{\boldmath{Higgs boson pair production at the LHC in the $b \bar{b} W^+ W^-$ channel}}

\author{Andreas Papaefstathiou$^b$, Li Lin Yang$^{a,b}$, Jos\'e Zurita$^b$}
\affiliation{
  \mbox{$^a$Department of Physics and State Key Laboratory of Nuclear Physics and Technology, Peking University, Beijing 100871, China}
  \\
  \mbox{$^b$Institut f\"ur Theoretische Physik, Universit\"at Z\"urich, 8057 Z\"urich, Switzerland}
}

\begin{abstract}
  We consider Higgs boson pair production at the LHC in the $b \bar{b} W^+ W^-$ channel, with subsequent decay of the $W^+W^-$ pair into $\ell \nu j j$. Employing jet substructure and event reconstruction techniques, we show that strong evidence for this channel can be found at the 14~TeV LHC with 600~fb$^{-1}$ of integrated luminosity, thus improving the current reach for the production of Higgs boson pairs. This measurement will allow to probe the trilinear Higgs boson coupling $\lambda$.
\end{abstract}


\maketitle

\noindent \textbf{Introduction.} One of the aims of the LHC is to search for the agent of electroweak symmetry breaking (EWSB), which in its minimal form is the Standard Model (SM) Higgs boson ($h$). Recently, both the ATLAS and the CMS collaborations have observed a new state with a mass of about 125~GeV, whose properties are in substantial agreement with the SM Higgs boson~\cite{LHCHiggs}. The quest for understanding the mechanism behind EWSB does not end with the discovery of the Higgs boson. It is crucial to test the Higgs boson potential to its full extent, measuring the couplings of the Higgs boson to gauge bosons and matter fields, and also to probe its self interactions. After EWSB, the Higgs potential can be written as $V(h) = m_h^2h^2 / 2 + \lambda v h^3 + \tilde{\lambda} h^4 / 4$. In the SM, $\lambda=\tilde{\lambda}= (m_h^2 / 2 v^2) \approx 0.13$ for $m_h$=125 GeV. With an extended Higgs sector, as is common in many new physics models beyond the SM, these couplings will deviate from the SM values. Therefore, measuring these two couplings is very important to reveal the true nature of the Higgs boson. At the LHC, the quartic coupling $\tilde{\lambda}$ may be probed via triple Higgs boson production. However, its tiny cross section \cite{Plehn:2005nk} makes it very difficult, if not impossible, to do so. On the other hand, the trilinear coupling $\lambda$ can be measured with Higgs boson pair production, $pp \to hh$, which may be discovered at a large luminosity phase of the LHC. In the following we will focus on that possibility.

The discovery potential of Higgs boson pair production at the LHC has been studied in \cite{Baur:2003gp, Dolan:2012rv}. Ref.~\cite{Baur:2003gp} concentrated on the decay channels $hh \to b\bar{b}\gamma\gamma, b\bar{b}\mu^+\mu^-$, finding that with 600~fb$^{-1}$ one expects 6 signal and 11 background events, giving a significance of about 1.5$\sigma$. In the recent years, jet substructure has been shown to be very important when dealing with hadronic decays of heavy particles \cite{Altheimer:2012mn}. In the $h \to b \bar{b}$ case, this was put forward in the seminal paper by Butterworth, Davison, Rubin and Salam (BDRS)~\cite{Butterworth:2008iy} in the context of $Wh$ and $Zh$ production, which were previously considered as challenging to probe at the LHC. With the subjet techniques, BDRS have shown that this can become a very promising discovery channel for the Higgs boson. Ref.~\cite{Dolan:2012rv} employed these new techniques, and assuming good $\tau$ reconstruction efficiency ($\sim 80\%$), the authors claimed the $b\bar{b}\tau^+\tau^-$ channel as the most promising one, with 57 signal and 119 background events at 600~fb$^{-1}$.

In both \cite{Baur:2003gp} and \cite{Dolan:2012rv}, the $hh \to b\bar{b}W^+W^- \to b\bar{b} \ell\nu jj$ channel was considered less promising, due to the large $t\bar{t}$ background. In this Letter, we apply the BDRS techniques to this final state in conjunction with event reconstruction using mass-shell constraints, assuming that the Higgs boson mass is well-measured. We show that in the highly boosted regime, the reconstruction of both Higgs bosons present in the event allows us to distinguish the signal and background, thereby turning this channel into a potentially significant contribution in the discovery of Higgs boson pair production.

\noindent \textbf{Higgs pair production and decay.} The main production mechanism for Higgs boson pairs at the LHC is gluon fusion, which was studied at leading order (LO) in quantum chromodynamics (QCD) in \cite{Glover:1987nx, Plehn:1996wb}. Other production modes such as $qq \to qqhh$, $Vhh$, $t \bar{t} hh$ are a factor of 10-30 smaller~\cite{Djouadi:1999rca, Gianotti:2002xx}, and therefore we do not consider them in the rest of our analysis.

\begin{figure}[t]
  \centering
  \begin{tabular}{cc}
    \includegraphics[width=0.5\linewidth]{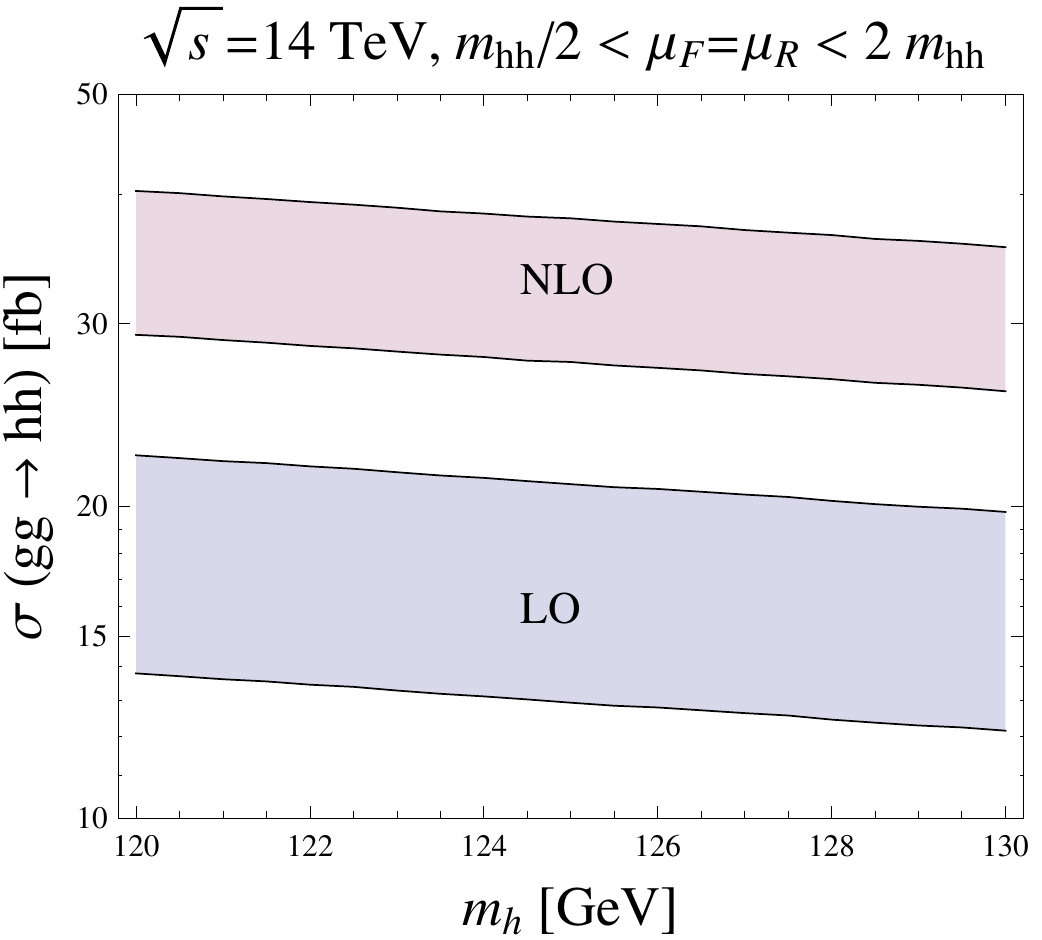}
    &
    \includegraphics[width=0.5\linewidth]{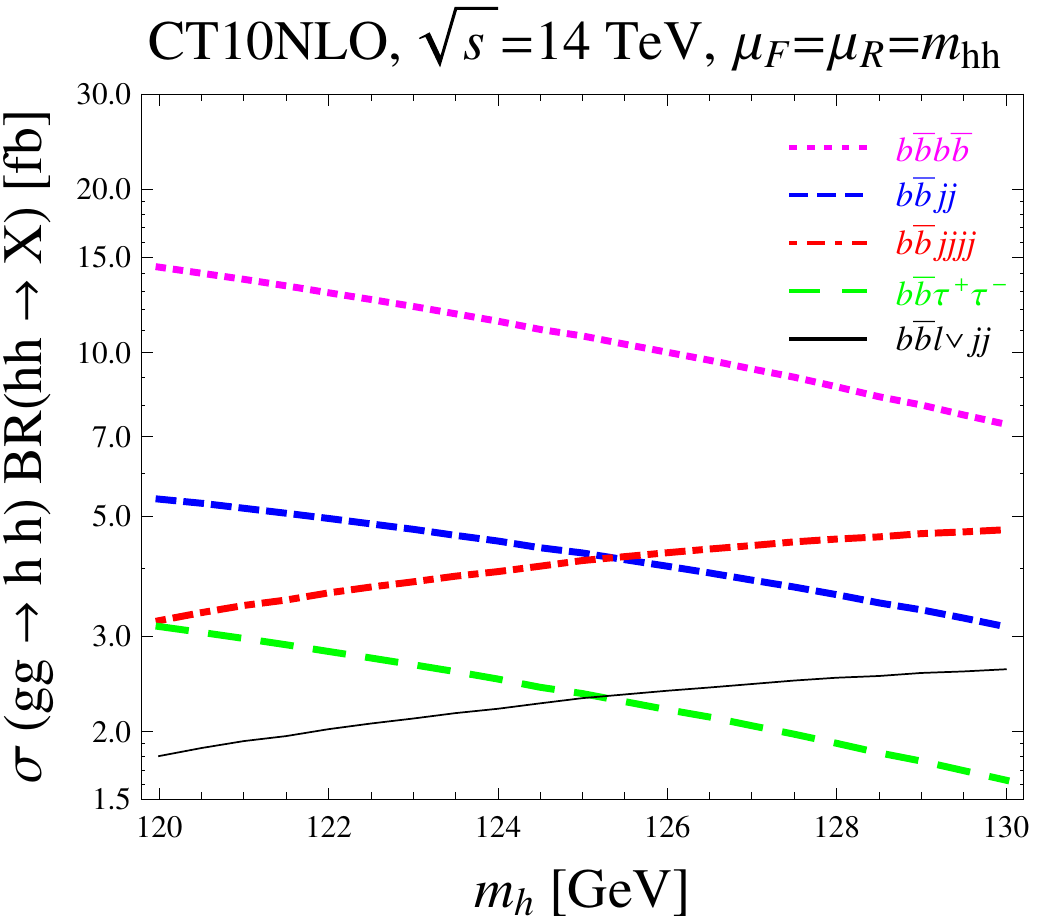}
    \\
    (a) & (b)
  \end{tabular}
  \vspace{-1ex}
  \caption{(a) Scale variation of gluon fusion cross section for Higgs boson pair production, at LO and NLO. (b) cross sections times branching ratios at the 14 TeV LHC, for Higgs boson pair production. We show only the dominant decay modes: $b \bar{b} b \bar{b}$ (dots), $b \bar{b} jj$ (short dashes), $b \bar{b} b jjjj$ (dot-dashes), $b \bar{b} \tau^+ \tau^-$ (long dashes) and $b \bar{b} l \nu jj$ (solid). Note that the four main decay modes are fully hadronic.}
  \label{fig:hpairLHC}
\end{figure}

We employ the code \texttt{HPAIR}~\cite{hpair} to compute the production cross section, which implements the next-to-leading order (NLO) QCD corrections obtained in the heavy top quark limit~\cite{Dawson:1998py}. We have modified the public version of \texttt{HPAIR} in order to use the up-to-date parton distribution functions (PDFs) present in the LHAPDF library~\cite{Whalley:2005nh}. For the LO and NLO cross sections, we employ CTEQ6L1 and CT10~\cite{Lai:2010vv} PDF sets with the corresponding values of $\alpha_s$, respectively. We adopt the pole masses for the top and bottom quarks to be $m_t=174.0$~GeV and $m_b=4.5$~GeV. For a 125~GeV Higgs boson, we have obtained an NLO cross section of $32.3^{+5.6}_{-4.7}$~fb, where the uncertainty reflects the variation of the renormalization and factorization scales $\mu_r=\mu_f$ around the central value $\mu_0$ by a factor of 2, with $\mu_0$ being the Higgs boson pair invariant mass. In the left panel of Fig.~\ref{fig:hpairLHC} we show the scale variation of the production cross sections at LO and NLO. One can observe that there is a large K-factor ($\sim 2$) on the cross section, and that the scale uncertainty is still high (about 20\%). Either an NNLO computation or performing QCD resummation could help reducing the scale uncertainty.

For the branching ratios, the values of \cite{Dittmaier:2011ti, Kniehl:2012rz} were used. In the mass range $(120, 130)$~GeV, the Higgs boson decay modes with the largest branching fractions are $h \rightarrow b\bar{b}$ and $h\rightarrow W^+W^-$. The most probable decay mode for a pair of Higgs bosons is $hh \to b\bar{b} b\bar{b}$. This mode is challenging to search for, mostly due to the fact that it is difficult to trigger on, and that it competes against the QCD multi-jet backgrounds that possess overwhelmingly large cross sections. In general, QCD backgrounds can be suppressed with the existence of leptons and missing energy. We plot in the right panel of Fig.~\ref{fig:hpairLHC} the total rates for the five most important channels at the 14 TeV LHC, with the Higgs mass in the 120-130 GeV range. As can be seen, the first four channels are purely hadronic. The most important channel that contains leptons and missing energy is $b\bar{b}W^+W^-$ with $W^+W^- \rightarrow \ell \nu j j$, where $\ell$ is either an electron or a muon and $j$ refers to light jets. For a 125 GeV Higgs boson, the branching ratio for this mode is $\sim 7.25\%$ \cite{Kniehl:2012rz}, and the total rate is $\sim 2.34$~fb.

\noindent \textbf{Event generation and analysis.} We now describe our analysis strategy for the $b\bar{b} \ell\nu jj$ channel. We will focus on a `mid-term' integrated luminosity of 600~fb$^{-1}$ for the LHC at a center-of-mass energy of 14~TeV. The largest background for this final state is $t\bar{t}$ production with semi-leptonic decay of the top pair. This background is the most challenging one: not only it has a large total rate ($\sim 240$~pb), but also possesses a mass scale, given by the top mass ($\sim 175$~GeV). The second important background is $W(\to \ell \nu)b\bar{b}$+jets, with a total rate of $\sim 2.17$~pb. Other QCD multi-jets production associated with a $W$ boson can enter, with two light jets misidentified as coming from $b$-quarks. Backgrounds originating from associated production of a single Higgs boson can also be present: $h(\rightarrow WW)b\bar{b}$, $h(\rightarrow b\bar{b})WW$ and $h$+jets where the jets are miss-identified.

Parton-level events of the $hh$ signal, with the Higgs boson mass set to 125~GeV at the 14~TeV LHC have been generated using a custom \texttt{MadGraph}~5 model~\cite{Alwall:2011uj, RikPriv}, which includes the full top quark mass effects in the relevant box and triangle diagrams. The factorization and renormalization scales are set to $\mu_F = \mu_R = 125$~GeV, and we checked that other scale choices do not substantially alter the conclusions of our analysis. The decays of the Higgs bosons are performed in \texttt{HERWIG++}~\cite{Bahr:2008pv, Arnold:2012fq}, and the total rate is normalized to the NLO value of 2.34~fb. The $t\bar{t}$ background is generated using \texttt{HERWIG++} with subsequent semi-leptonic decay, whose cross section is normalized to the approximate NNLO value (times branching ratio) of 240~pb~\cite{Ahrens:2011px}. Parton-level events for other backgrounds are generated using \texttt{ALPGEN}~\cite{Mangano:2002ea}, where the transverse momenta of light partons or $b$-quarks were constrained to be $p_T > 30$~GeV and their separation satisfies $\Delta R = \sqrt{(\Delta y)^2+(\Delta \phi)^2} > 0.35$, with $y$ and $\phi$ being the rapidity and azimuthal angle, respectively. The parton-level events are then showered and hadronized via \texttt{HERWIG++}. Whenever applicable, MLM-matching~\cite{Mangano:2002ea} as implemented in \texttt{HERWIG++}~\cite{Arnold:2012fq} is used to avoid double-counting in certain regions of phase space.

The hadron-level particles satisfying $p_T > 0.1$~GeV and $|\eta| < 5$ are clustered into jets with the Cambridge-Aachen algorithm using \texttt{FastJet}~\cite{Cacciari:2011ma}, with a radius parameter $R=1.4$. We then pick those jets with $p_T > 40$~GeV, which results in what we call `fat' jets. For a given fat jet $j$, we then examine its subjets $j_1$ and $j_2$ (with $m_{j_1} > m_{j_2}$) following the BDRS~\cite{Butterworth:2008iy} procedure. We ask for a significant mass drop $m_{j_1} < \mu m_j$ with $\mu = 0.667$, and require that the splitting is not too asymmetric by imposing $\min(p_{T,j_1}^2, p_{T,j_2}^2) \Delta R^2_{j_1,j_2} / m_j^2 > 0.09$. We also apply a `filtering' procedure similar to that applied by BDRS: resolving the fat jets on a finer angular scale $R_{\mathrm{filt}} < R_{j_1,j_2}$ and taking the three hardest objects (subjets) that appear, where we choose $R_{\mathrm{filt}} = \min (0.35, R_{j_1,j_2}/2)$. This provides versatility to the analysis against the effects of extra radiation, particularly the underlying event. In the present study we do not consider the effects of the detector resolution, which of course have to be included in a detailed experimental study.

We look for events containing at least two filtered fat jets satisfying the mass drop condition. We then impose the following conditions:
\begin{enumerate}
\item{Exactly one isolated lepton with $p_{T,\ell} > 10$~GeV and $|\eta| < 2.5$, where isolation means that the scalar sum of the transverse momenta of the visible particles lying inside a cone of radius $R = 0.15$ around the lepton is less than $0.1 \times p_{T,\ell}$.}
\item{Missing transverse energy $\met > 10$~GeV.}
\item{At least one fat jet with its two leading subjets $b$-tagged, which satisfies $|\eta| < 2.5$, $p_T > 180$~GeV and $m \in [115-135]$~GeV. Among these we take the one with highest $p_T$ as the $h \to b\bar{b}$ candidate and refer to it as $h_1$. The system of the two $b$-tagged subjets is referred to as $b\bar{b}$.}
\item{A second fat jet with $p_T > 40$~GeV and $m > 5$~GeV, which, together with the lepton and $\met$, can reconstruct the $W$-decaying Higgs boson ($h_2$). This jet will be considered as candidate for the hadronically decaying $W$ boson, and will be referred to as $W_h$.}
\end{enumerate}
In the above, $b$-tagging is implemented in the event generators by keeping the lightest $B$-hadrons stable. Throughout this work we assume a $b$-tagging efficiency of 70\%. The reconstruction of the $W$-decaying Higgs boson is achieved by solving the set of equations $m_h^2 = (p_\ell + p_\nu + p_{W_h})^2$ and $p_\nu^2 = 0$, where the transverse components of $p_\nu$ are identified with those of the missing transverse momentum. Here we assume that the mass of the Higgs boson will already have been measured to a reasonable accuracy. Note that since the equations are quadratic, there are two solutions for the $z$-component of momentum of the neutrino. It is, however, not possible to decide which is the correct one and we therefore do not use this information in our analysis. Here we reject events giving complex solutions, although one may adopt some imaginary part `tolerance' to accommodate the smearing of the momenta by detector effects~\cite{Gripaios:2010hv}.

\begin{table}[t!]
  \begin{center}
    \begin{tabular}{|l|l|l|}
      \hline
      Process & $\sigma_{\mathrm{initial}}$ (fb) &
      $\sigma_{\mathrm{\textbf{basic}}}$ (fb)
      \\ \hline 
      $hh \rightarrow  b\bar{b} \ell \nu jj$ & 2.34 &  0.134
      \\ \hline \hline
      $t\bar{t} \rightarrow b\bar{b} \ell \nu jj$ & $240 \times 10^3$
      & 15.5
      \\
      $W(\rightarrow \ell \nu)b\bar{b}$+jets & $2.17 \times 10^3$ &
      0.97
      \\ \hline \hline
      $W(\rightarrow \ell \nu)$+jets &  $2.636 \times 10^6$ &
      $\mathcal{O}(0.01)$
      \\ \hline
      $h(\rightarrow \ell \nu jj)$+jets & $36.11$ &
      $\mathcal{O}(0.0001)$
      \\
      $h(\rightarrow \ell \nu jj)b\bar{b}$ & $6.22$ &
      $\mathcal{O}(0.001)$
      \\
      $h(\rightarrow b\bar{b}) + WW(\rightarrow \ell\nu jj)$ &
      0.0252 & -
      \\
      \hline
    \end{tabular}
  \end{center}
  \caption{Cross sections for the signal and backgrounds before (second column) and after (third column) the `basic' cuts. For the irreducible backgrounds where true $b$-quarks are not present, a miss-$b$-tagging probability of 1\% for light jets are included. The MLM-matching is applied to the $Wb\bar{b}$+jets, $W$+jets and $h$+jets processes.}
  \label{tb:bcuts}
\end{table}

The conditions described above will be referred to as the `basic' cuts, and already provide strong rejection against backgrounds. Table~\ref{tb:bcuts} shows the starting cross sections for the processes considered as well as the resulting cross sections after the `basic' cuts. Among the irreducible backgrounds where the final states are exactly the same as our signal, the important ones are $t\bar{t}$ and $Wb\bar{b}$+jets, which we will further analyze, while the $hb\bar{b}$ and $hWW$ processes are negligible. The $W$+jets background requires two miss-$b$-tagged light jets to fake our signal. We estimate the rejection factor as follows: for the $W$+jets inclusive sample, we pick the hardest filtered fat jet and, assuming that its two hardest filtered subjets are miss-$b$-tagged, we apply the `basic' cuts to the event. We multiply the resultant cross section by the light jet rejection factor ($10^{-4}$, assuming the light jet miss-$b$-tag probability to be 1\%) for two jets. The $h+$jets background also requires miss-$b$-tags, for which we work in the same way as with the $W$+jets. These reducible backgrounds are found to be irrelevant after the `basic' cuts.

\begin{figure}[t]
  \centering
  \begin{tabular}{cc}
    \includegraphics[width=0.5\linewidth]{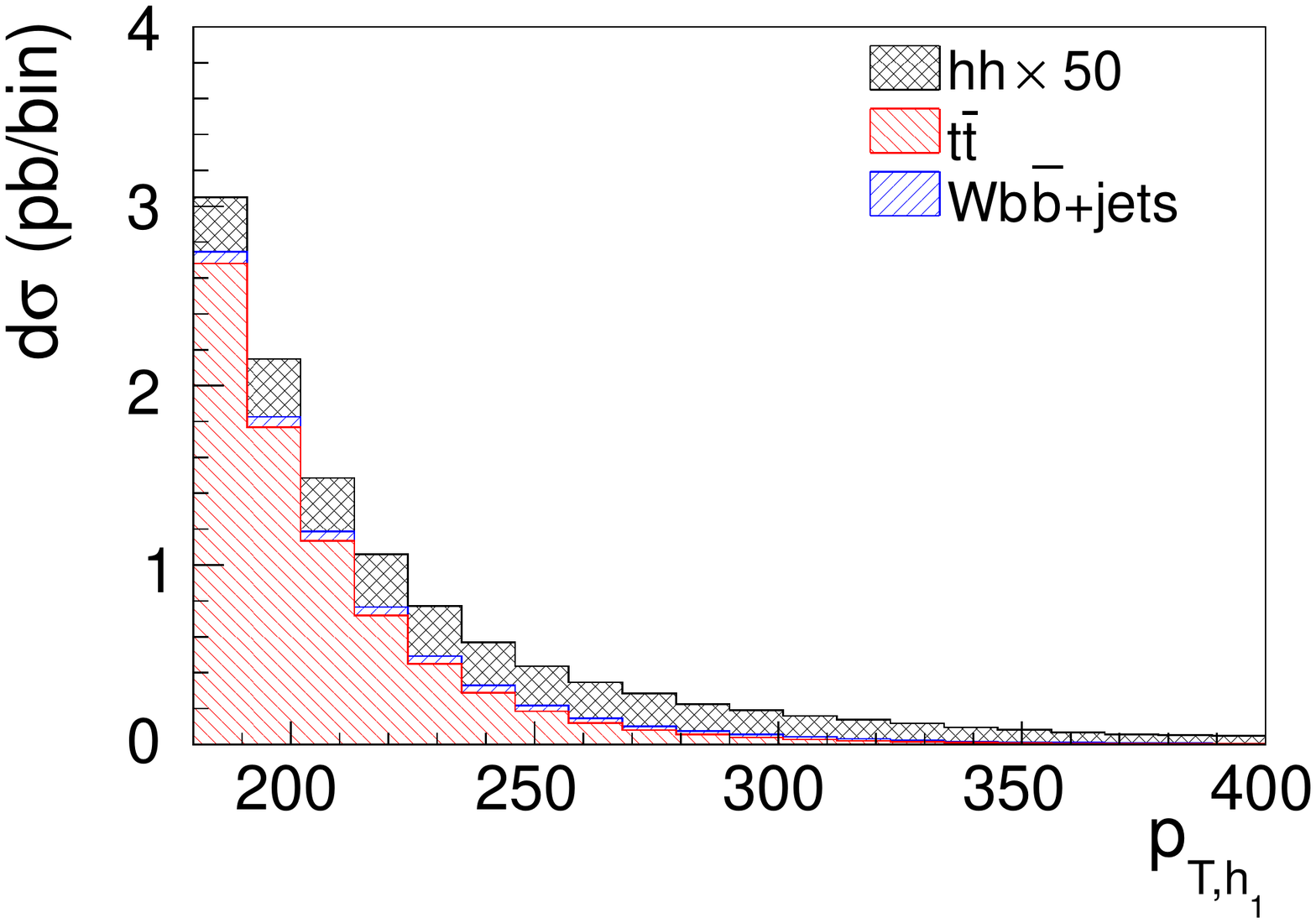}
    &
    \includegraphics[width=0.5\linewidth]{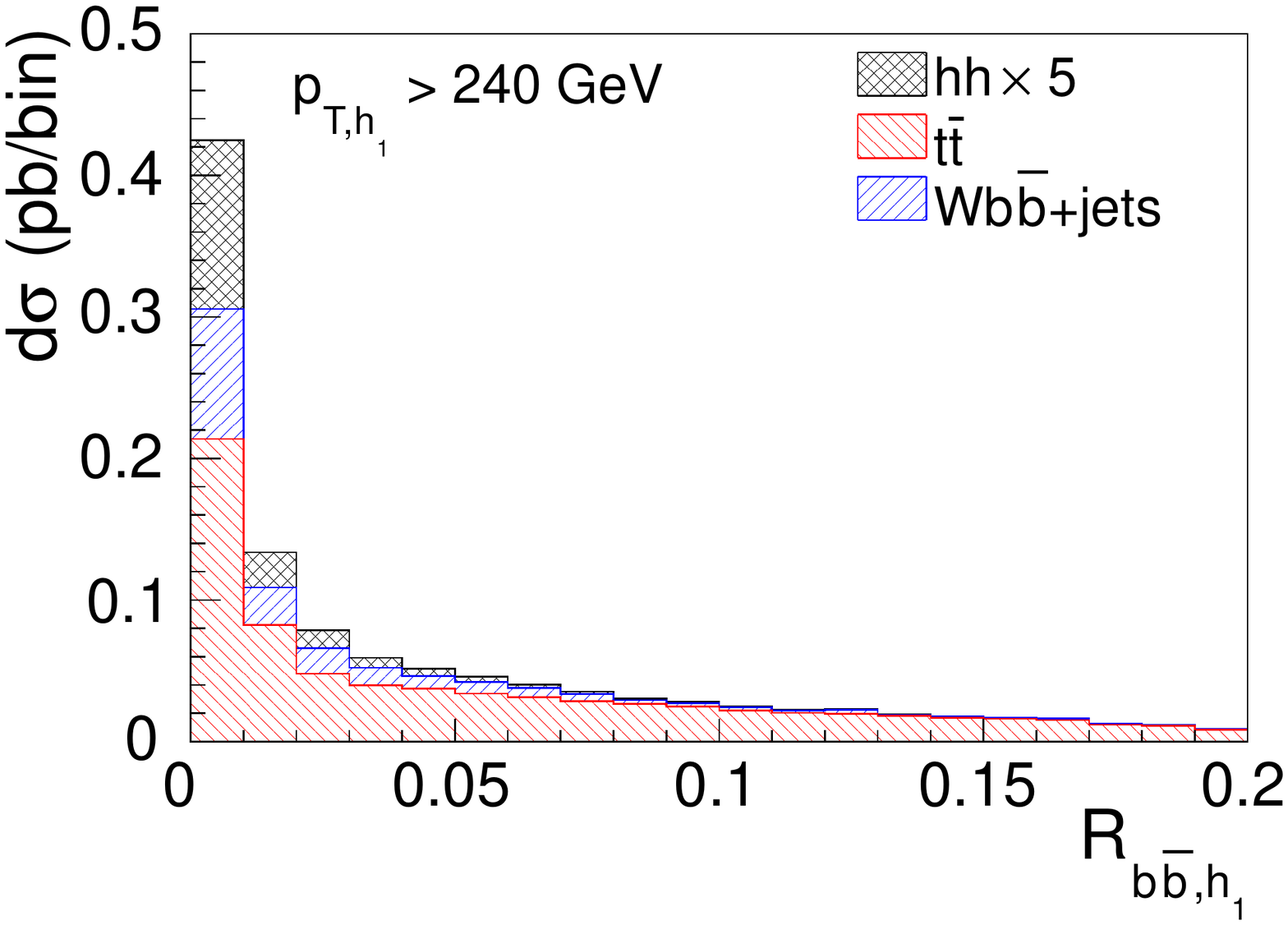}
    \vspace{-2ex}
    \\
    (a) & (b)
    \vspace{1ex}
    \\
    \includegraphics[width=0.5\linewidth]{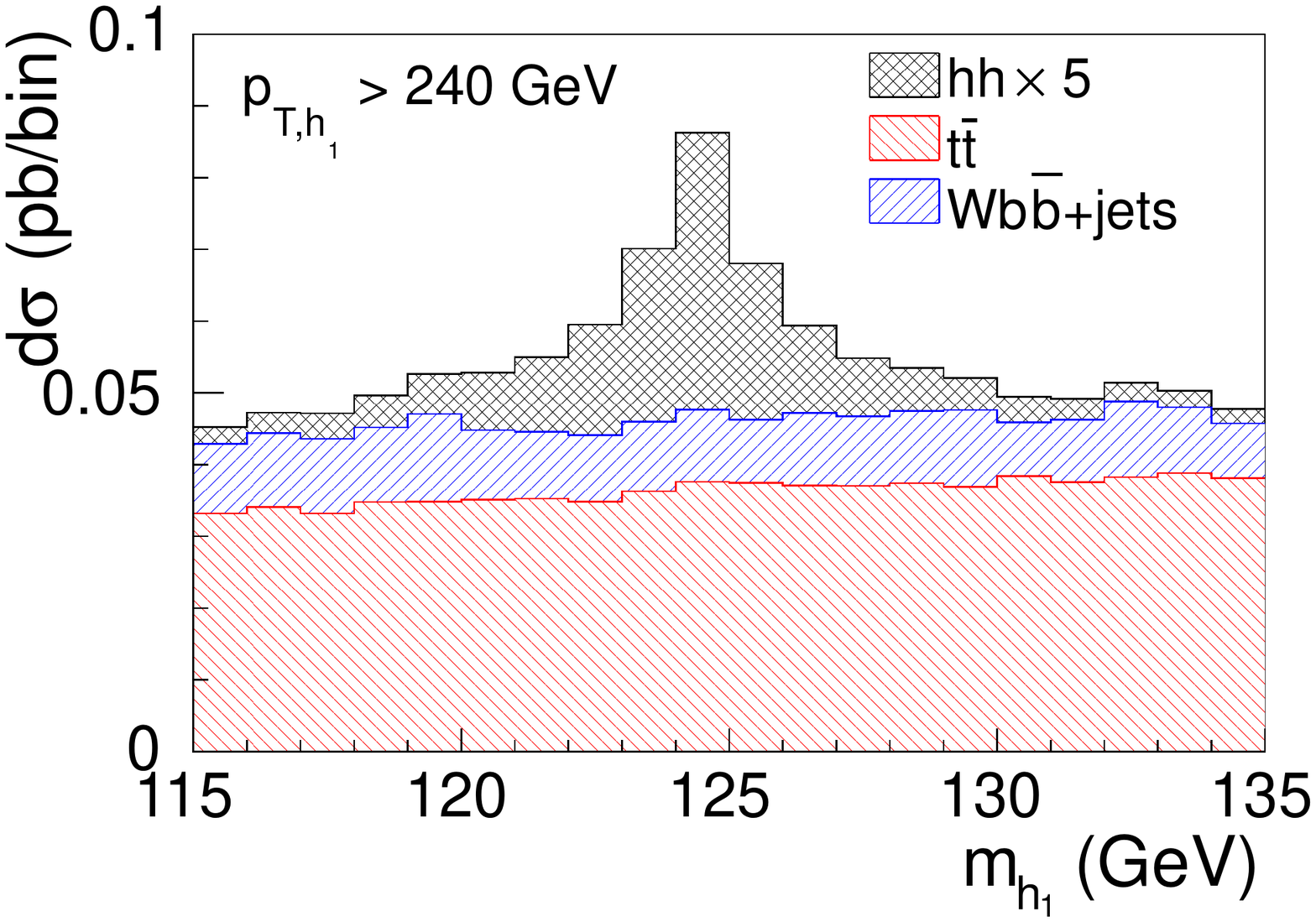}
    &
    \includegraphics[width=0.5\linewidth]{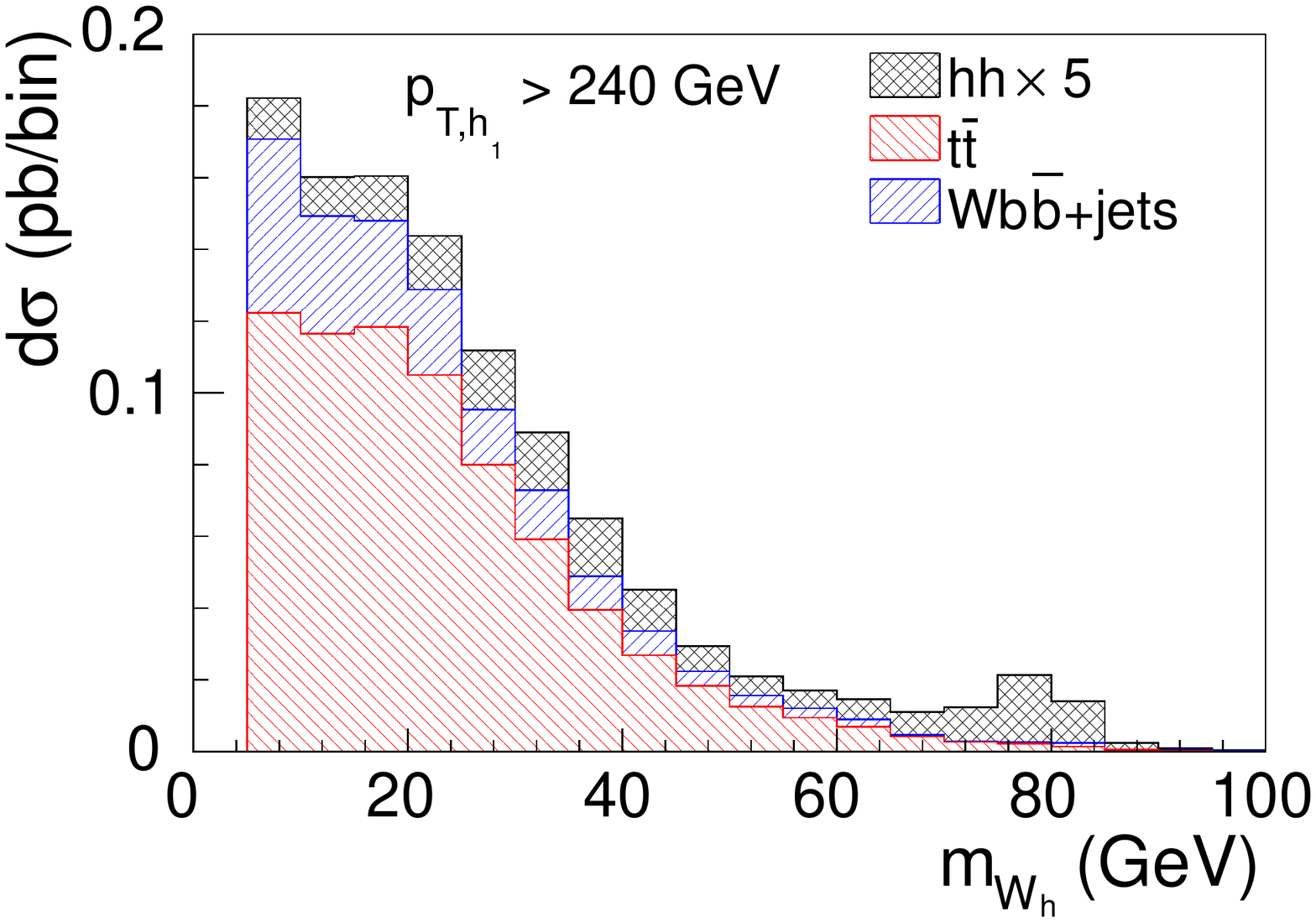}
    \vspace{-2ex}
    \\
    (c) & (d)
  \end{tabular}
  \vspace{-1ex}
  \caption{Distributions for signal and backgrounds of (a) $p_{T,h_1}$ after the basic cuts; and (b) $R_{b\bar{b},h_1}$, (c) $m_{h_1}$, (d) $m_{W_h}$ after the basic cuts and $p_{T,h_1} > 240$~GeV.}
  \label{fig:smart}
\end{figure}

We investigate in further detail the $hh$ signal versus the $t\bar{t}$ and $Wb\bar{b}$+jets backgrounds, going beyond the `basic' cuts. We show the signal ($S$) and background ($B$) distributions to demonstrate the set of cuts that provides a high significance, while retaining a reasonable number of signal events in order to keep the statistical error under control. We show in Fig.~\ref{fig:smart}(a) the $p_{T,h_1}$ distributions, where we see that the signal tends to have a larger $p_T$ for the Higgs candidate. We therefore impose a harder cut $p_{T,h_1} > 240$~GeV and subsequently consider the (b) $R_{b\bar{b},h_1}$ (distance between the $h_1$ fat jet and the $b\bar{b}$ sub-system), (c) $m_{h_1}$ and (d) $m_{W_h}$ distributions. One can observe that significant background rejection can be obtained by selecting $m_{W_h}$ around the $W$ boson mass $m_W$, requiring that the $b$ and $\bar{b}$ subjets are more symmetrically distributed in the fat jet $h_1$ by choosing a small $R_{b\bar{b},h_1}$, and imposing a mass window for $m_{h_1}$ around the true Higgs mass $m_h$. We choose $m_{W_h} > 65$ GeV, $m_{h_1} \in [120-130]$ GeV and $R_{b\bar{b},h_1} < 0.06$. Using these simple cuts, we obtain about 4.6 signal and 2.6 background events at 600~fb$^{-1}$, thus getting $S / \sqrt{S+B} \sim 1.7$, and a significance of $2.2\sigma$. To gain more discriminating power, we explored in more detail the kinematic distributions of the various objects. While a cut-based method is possible (we managed to achieve $2.5\sigma$ with $S \approx 4$ and $B \approx 1$), we performed a more dedicated multivariate analysis for that purpose. To this end we employ the boosted decision tree (BDT) method \cite{Roe:2004na} implemented in the \texttt{ROOT} \texttt{TMVA} package~\cite{Hocker:2007ht}. In addition to our previous set of variables, we add the following: $p_{T,h_2}$, $p_{T,W_h}$, $p_{T,h_1h_2}$, $R_{h_1,W_h}$, $M_{T,\ell\nu}$, $\Delta\phi_{\ell,\nu}$, $\Delta\phi_{W_l,W_h}$, where $W_l$ refers to the leptonically decaying $W$ boson, and the transverse mass of the lepton and neutrino system is defined as $M_{T,\ell\nu}^2 \equiv (E_{T,\ell}+E_{T,\nu})^2 - (\vec{p}_{T,\ell}+\vec{p}_{T,\nu})^2$.

\begin{figure}[t]
  \centering
  \begin{tabular}{cc}
    \includegraphics[width=0.5\linewidth]{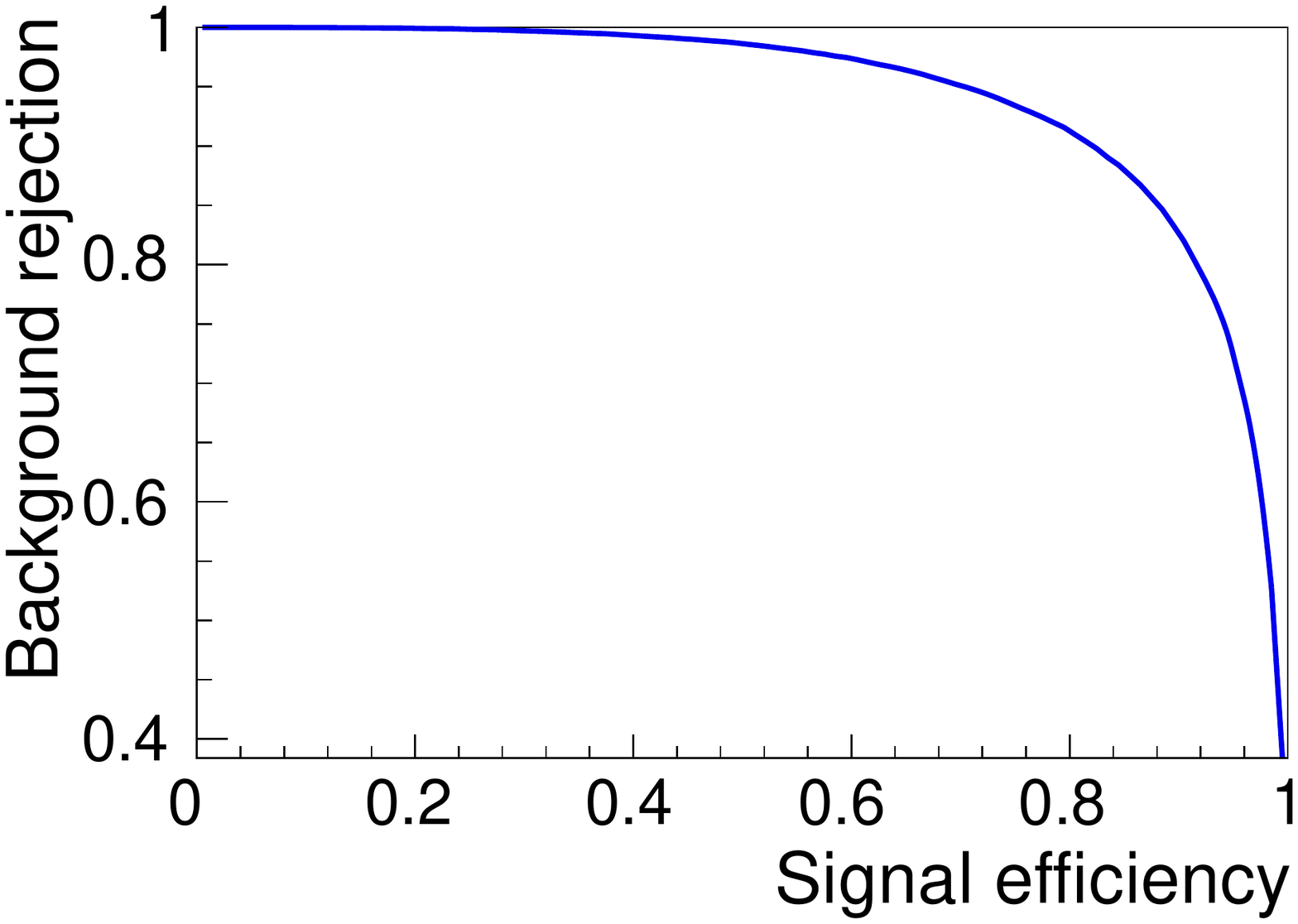}
    &
    \includegraphics[width=0.5\linewidth]{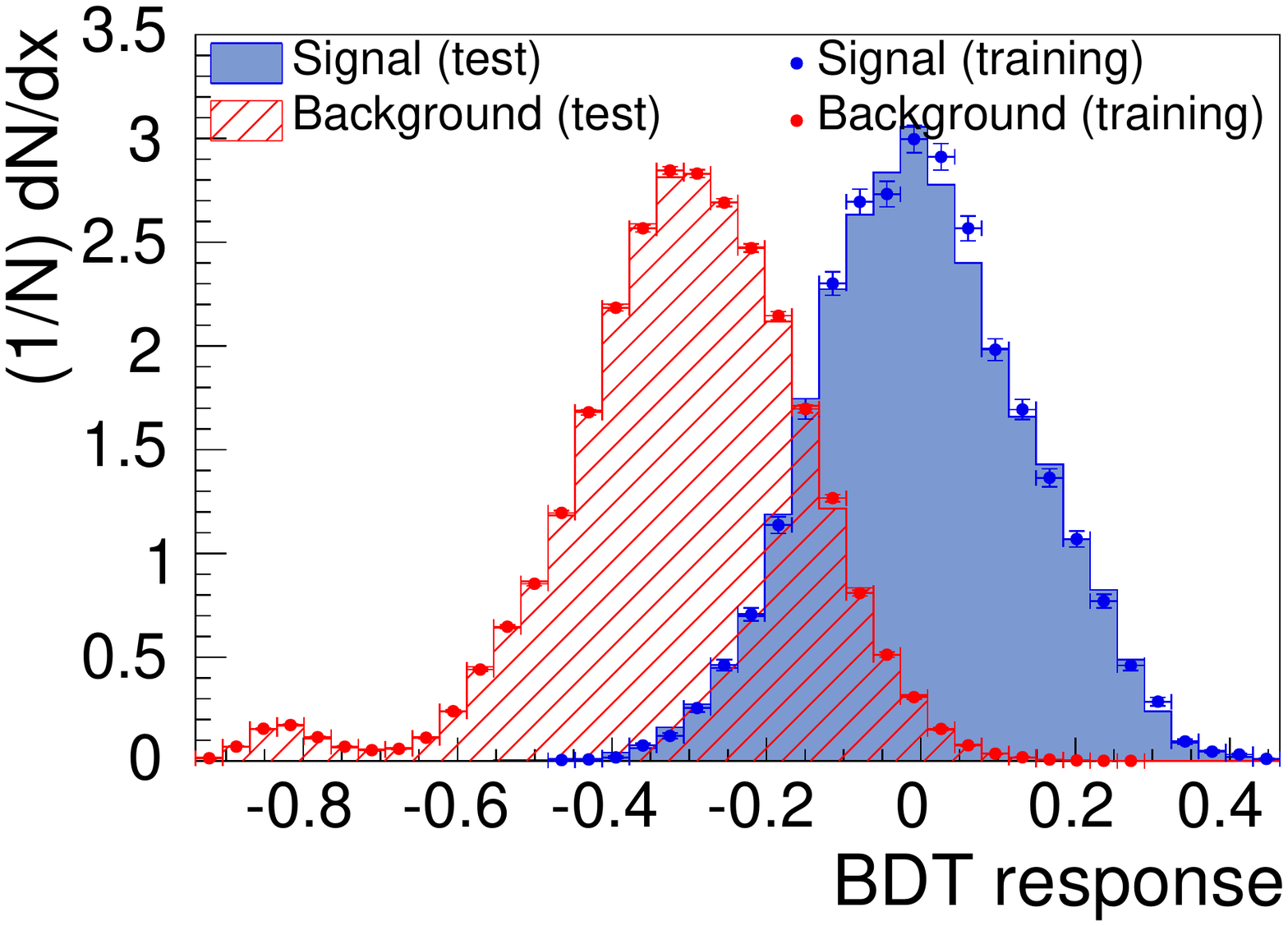}
  \end{tabular}
  \vspace{-1ex}
  \caption{Outputs of BDT analysis. Left: background rejection vs. signal efficiency. Right: normalized signal and background distributions against BDT response.}
  \label{fig:bdt}
\end{figure}

We trained 1000 decision trees, from which the outputs are shown in Fig.~\ref{fig:bdt}, where we can see that one can obtain good discrimination between signal and background. We find that when cutting at a value of around 0.1, we can obtain $S / \sqrt{S+B} \sim 2.4$ and a significance of $3.1\sigma$, with $S \approx 9$ and $B \approx 6$. We have checked that the inclusion of underlying event for the signal sample does not bring down the significance substantially. Further improvement can be obtained if one consider the tauonic decays of the $W$ bosons in both signal and background. Assuming a $\tau$ reconstruction efficiency of $\sim 70 \%$, one can obtain an increased significance of $3.6~(3.0)$ using the BDT (cut-based) analysis.

\noindent \textbf{Conclusions.} We have studied the prospects of detecting Higgs boson pair production at the 14 TeV LHC in the $b \bar{b} \ell \nu j j$ channel, where $\ell$ is either a muon or an electron. Our analysis is based on exploiting jet substructure techniques to identify the $h \to b\bar{b}$ decay for a Higgs boson in the boosted regime as a fat jet, and also event reconstruction for the $h \to W^+W^-$ decay. In spite of the very tiny initial signal to background ratio, we have identified a few useful kinematic variables that allow to discriminate signal from background. By cutting on these variables one can achieve an $\mathcal{O}(1)$ signal to background ratio, although retaining only a few handful of events for 600/fb. Further increase in the sensitivity can be achieved by including several more variables into the analysis. Given that scenario, we turned to a multivariate boosted decision tree analysis, which allows to obtain a significance of about $3\sigma$ while retaining a larger number (about 10) of signal events. Furthermore, the significance can be enhanced if we consider tau leptons in the final state, allowing to obtain just under $4 \sigma$ of sensitivity. This channel will make an important contribution, in combination with the already studied $b \bar{b} \tau^+ \tau^-$ and $b\bar{b} \gamma \gamma$, final states, towards the discovery of Higgs pair production at the LHC, and measuring the trilinear self interaction.

\noindent \textbf{Acknowledgments.} We would like to thank Christoph Englert for useful discussions and Thomas Gehrmann for comments on the manuscript. We also thank Rikkert Frederix for providing a MadGraph model file for $hh$ production. This work was supported by the Swiss National Science Foundation under contract 200020-138206 and 200020-141360/1, and by the Research Executive Agency of the European Union under the Grant Agreement number PITN-GA-2010-264564 (LHCPhenoNet).

\end{document}